\documentclass[twocolumn,showpacs,amsmath,amssymb]{revtex4}
\usepackage[dvips]{graphicx}
\usepackage{dcolumn}
\usepackage{bm}

\begin{document}

\title{Model dependence of isospin sensitive observables at high densities}

\author{Wen-Mei Guo$^{1,2,3}$}
\author{Gao-Chan Yong$^{1,5}$}\thanks{Corresponding author: yonggaochan@impcas.ac.cn}
\author{Yongjia Wang$^{3,4}$}
\author{Qingfeng Li$^{3}$}
\author{Hongfei Zhang$^{4,5}$}
\author{Wei Zuo$^{1,5}$}
\affiliation{%
$^1${Institute of Modern Physics, Chinese Academy of Sciences, Lanzhou 730000, China}\\
$^2${University of Chinese Academy of Sciences, Beijing 100049, China}\\
$^3${School of Science, Huzhou Teachers College, Huzhou
313000, China}\\
$^4${School of Nuclear Science and Technology, Lanzhou University,
Lanzhou 730000, China}\\ $^5${State Key Laboratory of Theoretical
Physics, Institute of Theoretical Physics, Chinese Academy of
Sciences£¬Beijing, 100190}
}%


\begin{abstract}
Within two different frameworks of isospin-dependent transport
model, i.e., Boltzmann-Uehling-Uhlenbeck (IBUU04) and
Ultrarelativistic Quantum Molecular Dynamics (UrQMD) transport
models, sensitive probes of nuclear symmetry energy are simulated
and compared. It is shown that neutron to proton ratio of free
nucleons, $\pi^{-}/\pi^{+}$ ratio as well as isospin-sensitive
transverse and elliptic flows given by the two transport models
with their ``best settings'', all have obvious differences.
Discrepancy of numerical value of isospin-sensitive n/p ratio of
free nucleon from the two models mainly originates from different
symmetry potentials used and discrepancies of numerical value of
charged $\pi^{-}/\pi^{+}$ ratio and isospin-sensitive flows mainly
originate from different isospin-dependent nucleon-nucleon cross
sections. These demonstrations call for more detailed studies on
the model inputs (i.e., the density- and momentum-dependent
symmetry potential and the isospin-dependent nucleon-nucleon cross
section in medium) of isospin-dependent transport model used. The
studies of model dependence of isospin sensitive observables can
help nuclear physicists to pin down the density dependence of
nuclear symmetry energy through comparison between experiments and
theoretical simulations scientifically.
\end{abstract}

 \maketitle

\section{Introduction}

The equation of state of isospin asymmetric nuclear matter, i.e.
the density dependence of the symmetry energy especially its
high-density behavior, is still one of the open questions in
nuclear physics. In recent years, many nuclear physicists have
made great efforts to explore the density-dependent nuclear
symmetry energy, which has significant ramifications in
understanding the structure of rare isotopes, heavy-ion nuclear
reactions induced by radioactive beam
\cite{LiBA:1998,BaranV:2005,LiBA:2008}, and also in astrophysics
\cite{Sumiyoshi:1994,Lattimer:2004,Steiner:2005}. Around normal
density, the symmetry energy has been roughly constrained from,
e.g., studying isospin diffusion
\cite{MBT:2004,TXLiu:2007,ChenLW:2005,LiBA:2005} and isoscaling
\cite{DVShetty:2007} in heavy-ion reactions, the size of neutron
skin in heavy nuclei \cite{AWSteiner:2005}, and isotope dependence
of the giant monopole resonances in even-A Sn isotopes
\cite{LiT:2007}. However, there is a high degree of uncertainty
for constraints of nuclear symmetry energy in the high-density
areas. Different transport models give practically opposite
conclusions for the high-density dependence of the symmetry
energy, e.g., a very soft symmetry energy at the supra-saturation
density was indicated by fitting the FOPI data \cite{FOPI:2007}
based on the the isospin-dependent Boltzmann-Uehling-Uhlenbeck
(IBUU04) model \cite{XiaoZG:2009,xie2013} whereas Feng et al.
obtained a stiff result using the LQMD model \cite{FengZQ:2010}.
Other similar studies were also obtained in the QMD framework
\cite{russ11,cozma11}. This situation calls for the studies of
model dependence of probing the symmetry energy by using heavy-ion
collisions. It was exciting to note that a so-called
model-independent constraint of the high-density dependence of the
symmetry energy was recently obtained by Cozma \emph{et al.}
\cite{cozma2013}.

In fact, there are many factors affecting nuclear reaction
transport simulation, such as the initialization of colliding
nuclei, the nucleon-nucleon interaction potential, nucleon-nucleon
elastic and inelastic scattering cross sections, and the designs
of the framework of transport model codes
\cite{zhyx08,dds11,mdp,cozma11,epj,init}. So it is very necessary
to make a dialogue between different models, to see how large the
differences are on the values of isospin sensitive observables. To
give the model error estimation of observable actually involves
different transport calculations and one by one examining of the
effects of the uncertainties caused by different model inputs.
This is boring but important to read the experiment data
``correctly''. In this study, within the frameworks of isospin-%
dependent transport models Boltzmann-Uehling-Uhlenbeck (IBUU04)
and Ultrarelativistic Quantum Molecular Dynamics (UrQMD), we
investigated the model dependences of some frequently used
isospin-sensitive observables $\pi^{-}/\pi^{+}$ ratio and $n/p$
ratio of free nucleons and isospin-sensitive directed and elliptic
flows, which have been predicted to be sensitive to nuclear
symmetry energy \cite{BaranV:2005,LiBA:2008}.

\section{The transport models}

To simulate nuclear collisions, transport model that one
frequently utilized is the Boltzmann-Uehling-Uhlenbeck (BUU)
equation, which provides an approximate Wigner transform of the
one-body density matrix as its solution \cite{bertsch}. The BUU
transport model is usually used to describe one-body observable
although some afterburner can be added to predict many-body
correlation \cite{yongcluster}. The other frequently utilized
approaches is the Molecular Dynamics Model (QMD), which represents
the individual nucleons as Gaussian ``wave-packet'' with mean
values that move according to the Hamilton's equations
\cite{jorgaic}. QMD model has advantage over many-body correlation
and thus frequently used to predict cluster production in
heavy-ion collisions \cite{wangy13}. In the following discussions,
we use the isospin-dependent IBUU04 and UrQMD transport models to
discuss the model dependence of isospin sensitive observables at
high densities.

\subsection{The isospin-dependent BUU transport model}

In the used IBUU04 model, an isospin- and momentum-dependent
mean-field potential \cite{CBDas:2003} is used, i.e.,
\begin{eqnarray}
U(\rho,\delta,\textbf{p},\tau)&=&A_u(x)\frac{\rho_{\tau'}}{\rho_0}+A_l(x)\frac{\rho_{\tau}}{\rho_0}\nonumber\\
& &+B(\frac{\rho}{\rho_0})^{\sigma}(1-x\delta^2)-8x\tau\frac{B}{\sigma+1}\frac{\rho^{\sigma-1}}{\rho_0^\sigma}\nonumber\\
& &+\frac{2C_{\tau,\tau}}{\rho_0}\int d^3\,\textbf{p}'\frac{f_\tau(\textbf{r},\textbf{p}')}{1+(\textbf{p}-\textbf{p}')^2/\Lambda^2}\nonumber\\
& &+\frac{2C_{\tau,\tau'}}{\rho_0}\int
d^3\,\textbf{p}'\frac{f_{\tau'}(\textbf{r},\textbf{p}')}{1+(\textbf{p}-\textbf{p}')^2/\Lambda^2},
\label{buupotential}
\end{eqnarray}
where $\delta=(\rho_n-\rho_p)/(\rho_n+\rho_p)$ is the isospin
asymmetry, and $\rho_n$, $\rho_p$ are neutron ($\tau=1/2$) and
proton ($\tau=-1/2$) densities, respectively. Detailed parameter
settings can be found in Ref.~\cite{LiBA:2004}.
\begin{figure}[htb]
\centering\emph{}
\includegraphics[width=0.5\textwidth]{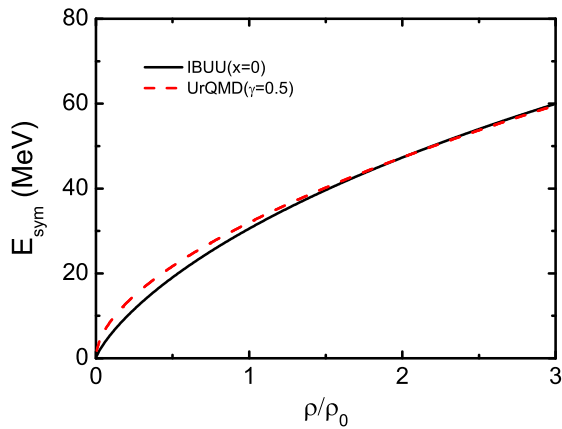}
\caption{Density dependent nuclear symmetry energies used in the
IBUU04 and the isospin-dependent UrQMD transport models.}
\label{fig:Fig1}
\end{figure}
\begin{figure*}[htb]
\centering\emph{}
\includegraphics[width=0.99\textwidth]{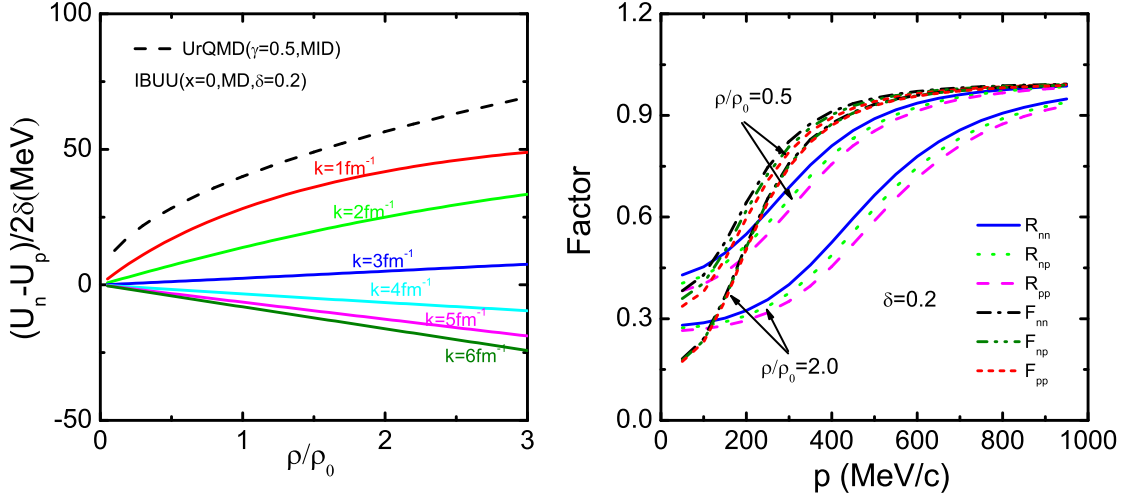}
\caption{Symmetry potentials and reduced medium correction factors
of $NN$ cross section used in the IBUU04 and the UrQMD transport
models. In the left panel, lines labelled by different momenta are
the symmetry potentials used in the IBUU04. In the right panel,
$F_{nn}, F_{np}, F_{pp}$ denote reduced factors used in the UrQMD,
and $R_{nn}, R_{np}, R_{pp}$ are reduced factors used in the
IBUU04 model.} \label{fig:Fig2}
\end{figure*}
The parameter $x$ is used for simulating different density
dependences of the symmetry energy $E_{sym}(\rho)$ predicted by
microscopic and phenomenological many-body approaches
\cite{AELD:2003}, but in our present work we just choose the
parameter $x=0$. Shown in Fig.~\ref{fig:Fig1} is the density
dependence of nuclear symmetry energy used in the IBUU04 transport
model and the following UrQMD transport model. From
Fig.~\ref{fig:Fig1} we can see that nuclear symmetry energies used
in the IBUU04 model (with $x=0$) and the UrQMD model (with
$\gamma=0.5$) are almost the same. However, the same
density-dependent symmetry energy does not mean the used symmetry
potential is also the same. The derived symmetry potential from
Eq.~\ref{buupotential} is shown in the left panel of
Fig.~\ref{fig:Fig2}, it is clearly seen that the symmetry
potential in the IBUU04 model is a density- and momentum-dependent
symmetry potential. For small momentum's nucleons, strength of the
symmetry potential increases with density. However, for large
momentum's nucleons strength of the symmetry potential decreases
with density. And we can clearly see that the symmetry potential
of the IBUU04 used is different from that used in the UrQMD model
for nucleons with nonzero momenta.

For the IBUU04 calculations, we also adopted an isospin-dependent
in-medium reduced $NN$ (nucleon-nucleon) elastic scattering cross
section, which originating from the scaling model according to
nucleon effective mass
\cite{LiBA:2005,JWNegele:1981,VRPand:1992,DPersram:2002}, i.e.,
based on the assumption that in-medium $NN$ scattering transition
matrix is the same as that in vacuum \cite{VRPand:1992}, the
elastic $NN$ scattering cross section in medium
$\sigma_{NN}^{medium}$ is reduced by a factor of
\begin{eqnarray}
R_{medium}(\rho,\delta,\textbf{p})&=&\sigma_{NN_{elastic}}^{medium}/\sigma_{NN_{elastic}}^{free}\nonumber\\
&=&(\mu_{NN}^*/\mu_{NN})^2,
\end{eqnarray}
where $\mu_{NN}$ and $\mu_{NN}^*$ are the reduced masses of the
colliding nucleon pair in free space and medium, respectively.
Momentum- and density-dependent reduced factors of $NN$ scattering
cross sections $R_{nn}, R_{np}, R_{pp}$ are shown in the right
panel of Fig.~\ref{fig:Fig2}. It is seen that the momentum- and
density-dependent reduced factor of $NN$ scattering cross section
used in the IBUU04 model demonstrates evident momentum- and
isospin-dependence. For in-medium $NN$ inelastic scattering cross
section, we use the experimental free space $NN$ inelastic
scattering cross section in the two transport models since the
medium effect of $NN$ inelastic scattering cross section is still
an open question.

\subsection{The isospin-dependent UrQMD transport model}

The UrQMD model is a microscopic model used to simulate
(ultra)relativistic heavy ion collisions and it has also been used
as a component of various hybrid transport approaches
\cite{urqmdcode}. In the isospin-dependent UrQMD model, the used
momentum-dependent potential was proposed by Bass \emph{et al.}
based on the mean field theory and expressed as \cite{Bass:1995}
\begin{eqnarray}
U_{md}=t_{md}\ln^2[1+a_{md}(\textbf{p}_i-\textbf{p}_j)^2]\frac{\rho_i}{\rho_0},
\end{eqnarray}
where $t_{md}=1.57\ MeV$ and $a_{md}=500\ c^2/GeV^2$. For the
symmetry potential energy density, we use the form of
\begin{eqnarray}
V_{sym}=(S_0-\frac{\varepsilon_F}{3})u^{\gamma}\delta^2
\end{eqnarray}
and the parameter settings are the same as in Ref.
\cite{LiQF:2011}. $S_0$ is the symmetry energy at normal nuclear
density $\rho_0$ and its value is about 30$\sim$ 36 MeV
\cite{DVretenar:2003,KPomorski:2003,AKlim:2007}. Here we choose
$S_0$= 32 MeV. $\varepsilon_F$ denotes the Fermi kinetic energy at
the normal nuclear density, which is approximately about 38 MeV.
$u=\frac{\rho}{\rho_0}$ is the reduced nuclear density, $\gamma$
is the strength parameter of the density dependence of symmetry
potential, and $\delta=(\rho_n-\rho_p)/(\rho_n+\rho_p)$ is the
isospin asymmetry. Here we adopt a soft ($\gamma=0.5$)
density-dependent symmetry potential, its corresponding
density-dependent symmetry energy is similar with $x= 0$ case used
in the IBUU04 model as shown in Fig.~\ref{fig:Fig1}. From the
symmetry potential energy density $V_{sym}$, one can get the
symmetry potential as a function of density as shown in the left
panel of Fig.~\ref{fig:Fig2}. It seems like the symmetry potential
used in the IBUU04 model at low ultimate momentum.

As for two-body scattering cross section in medium, it is somewhat
complicated than that used in the IBUU04 model. In-medium $NN$
elastic cross section is modified by nuclear medium according to
the QHD theory \cite{KCChou:1985,MaoGJ:1994,LiQF:2000}. In the
present work, the in-medium $NN$ elastic cross section
$\sigma_{NN_{elastic}}^{medium}$ comes from the free space elastic
scattering cross section $\sigma_{NN_{elastic}}^{free}$ multiplied
by a medium correction factor $F(u,\delta,p)$. It is formulated as
\begin{eqnarray}
\sigma_{NN_{elastic}}&=&F(u,\delta,p)\times\sigma_{NN_{elastic}}^{free}\\
&=& F_\delta^p\times F_u^p \times\sigma_{NN_{elastic}}^{free}.
\end{eqnarray}
The medium correction factor $F(u,\delta,p)$ is consist of the
momentum-dependent isospin-scalar density effect $F_u^p$ and the
momentum-dependent isospin-vector mass-splitting effect
$F_\delta^p$. Here the used non-relativistic neutron mass is
larger than that of proton in the neutron-rich medium, which is
consistent with the results of the Dirac-Brueckner-Hartree-Fock
(DBHF) theory or the extended Brueckner-Hartree-Fock (BHF) theory.
The isospin-dependent splitting effect on $NN$ elastic cross
section which represented by the $F_\delta$ factor has been
studied in Ref. \cite{LiQF:2006,LiQF:2010}. The momentum-dependent
reduced factors $F_u^p$ and $F_\delta^p$ are expressed in one
formula as
\begin{eqnarray}
F_{\delta,u}^p=\left\{%
\begin{array}{ll}
    1, & \hbox{$p_{NN}$$>$1 GeV/c;} \\
    \frac{F_{\delta,u}-1}{1+(p_{NN}/0.225)^3}+1, & \hbox{$p_{NN}$$\leq$1 GeV/c.} \\
\end{array}%
\right.
\end{eqnarray}
\begin{equation}
F_u=\frac{1}{6}+\frac{5}{6}e^{-3u},
\end{equation}
and isospin-dependent
\begin{eqnarray}
F_\delta=\left\{%
\begin{array}{ll}
    1-\frac{0.85}{1+3.25u}\delta, & \hbox{pp;} \\
    1+\frac{0.85}{1+3.25u}\delta, & \hbox{nn;} \\
    1, & \hbox{np.} \\
\end{array}%
\right.
\end{eqnarray}
Here $p_{NN}$ is the relative momentum of the two colliding
nucleons in the $NN$ center-of-mass system \cite{LiQF:2010}.
Momentum- and density-dependent reduced factors of $NN$ scattering
cross sections $F_{nn}, F_{np}, F_{pp}$ used in the UrQMD model
are shown in the right panel of Fig.~\ref{fig:Fig2}. It is seen
that the momentum- and density-dependent reduced factor of $NN$
scattering cross section used here demonstrates weak isospin
dependence. Compared with that used in the UrQMD model, the
reduced factor of $NN$ scattering cross section used in the IBUU04
model shows more density- and isospin-dependent.

\section{Results and discussions}

Generally speaking, the strength of symmetry potential (which
relates to the symmetry energy directly) is much weaker than the
strength of isoscalar potential. The other characteristic is that
the symmetry potential has opposite actions for neutrons and
protons. Therefore one always constructs observables of the
symmetry energy using differences or ratios of isospin multiplets
of baryons, mirror nuclei and mesons \cite{liyz06}. In the
following we mainly discuss the frequently used observables of
nuclear symmetry energy, i.e., nucleon or $\pi$ meson emissions
and nucleonic collective flow.

\subsection{nucleon and $\pi$ meson emissions}
\begin{figure*}[htb]
\centering\emph{}
\includegraphics[width=0.99\textwidth]{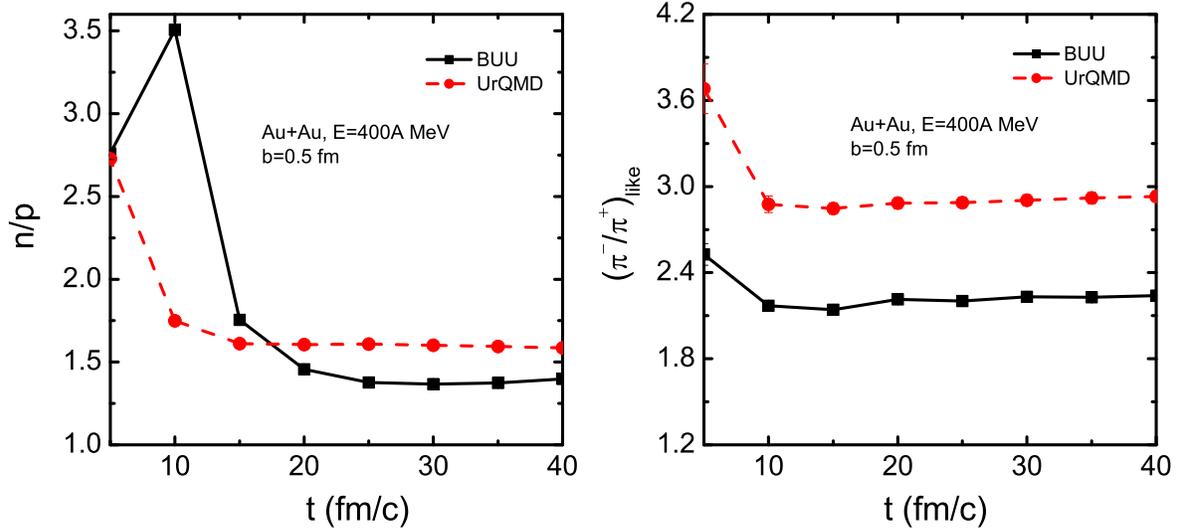}
\caption{Evolution of $n/p$ ratio of free nucleons and
$(\pi^{-}/\pi^{+})_{like}$ ratio in the central reaction of
$^{197}Au+^{197}Au$ at a beam energy of 400 MeV/nucleon. The black
solid line and red dashed line denote results of IBUU04 model
($t_{max}$= 40 fm/c) and UrQMD model ($t_{max}$= 150 fm/c),
respectively.} \label{fig:Fig3}
\end{figure*}
\begin{figure*}[htb]
\centering\emph{}
\includegraphics[width=0.99\textwidth]{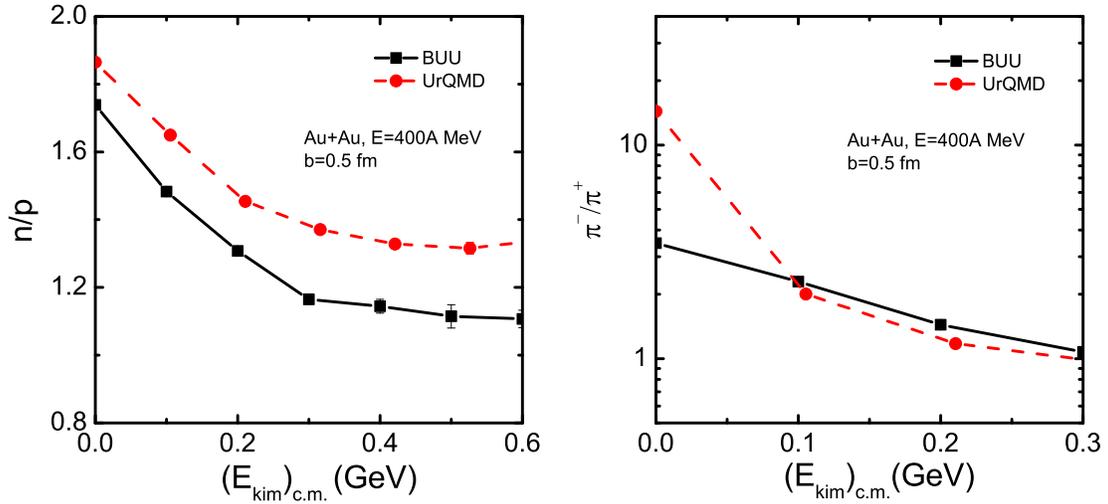}
\caption{$n/p$ ratio of free nucleons and $\pi^-/\pi^+$ ratio as a
function of kinetic energy in the central reaction of
$^{197}Au+^{197}Au$ at a beam energy of 400 MeV/nucleon. The black
solid line and red dashed line represent the results of IBUU04
model ($t_{max}$= 40 fm/c) and the UrQMD model ($t_{max}$= 150
fm/c), respectively.} \label{fig:Fig4}
\end{figure*}
Neutron to proton ratio of free nucleons as a probe of nuclear
symmetry energy in heavy-ion collisions was first proposed by Li
\emph{et al.} in 1997 \cite{liba97}. And $\pi^-/\pi^+$ ratio was
first proposed in 2002 as a probe of nuclear symmetry energy by Li
\cite{pionli2002}. Double neutron to proton ratio from isotopic
reaction systems was also proposed in 2006 as a probe of nuclear
symmetry energy \cite{liyz06} and t/$^{3}$He was proposed as a
similar probe as n/p in 2003 \cite{chen03}.  Later on double
$\pi^-/\pi^+$ ratio as a probe of the high-density behavior of the
nuclear symmetry energy was proposed in 2006 \cite{yong2006}.
After that, a lot of studies on such similar probes were carried
out in recent years
\cite{LiQF:2003,LiQF1:2006,DiToro:2010,YXZhang:2005,LiQF:2005,mayg12,maygprc12,gaoy12,
fengzq12}.

Shown in Fig.~\ref{fig:Fig3} are the evolutions of $n/p$ ratio of
free nucleons and $(\pi^{-}/\pi^{+})_{like}$ ratio in the central
reaction of $^{197}Au+^{197}Au$ at a beam energy of 400
MeV/nucleon simulated by isospin-dependent IBUU04 and UrQMD
models. With the dynamics of pion resonance productions and
decays, the $(\pi^-/\pi^+)_{like}$ ratio naturally becomes
$\pi^-/\pi^+$ ratio at final stage \cite{LiBA1:2005}. The large
difference of n/p of free nucleons emitted at the beginning from
the two models is due to different initializations of colliding
nuclei. Compared with the UrQMD's result, the small value of n/p
ration from the IBUU04 at final stage is caused by its smaller
value of the momentum-dependent symmetry potential than the
momentum-independent symmetry potential used in the UrQMD as shown
in the left panel of Fig.~\ref{fig:Fig2}. And the smaller value of
$\pi^-/\pi^+$ ratio from the IBUU04 calculation than that from the
UrQMD is due to its observable smaller $pp$ elastic cross section
than that of $nn$ (as shown in the right panel of
Fig.~\ref{fig:Fig2}, which causing relatively larger $pp$
in-elastic cross section than that of $nn$), thus relatively
larger number of $\pi^+$ mesons are produced than $\pi^-$, which
giving a smaller value of $\pi^-/\pi^+$ ratio than that of UrQMD.
In addition, different methods of constructing cluster
\cite{LiQingFeng:2009} and $NN$ inelastic cross section
\cite{LiQingFeng:2006} are also the reasons of model dependence.
From Fig.~\ref{fig:Fig3} we can clearly see that the $n/p$ ratio
of free nucleons and $\pi^-/\pi^+$ ratio given by UrQMD are,
respectively, $14.3\%$ and $30\%$ larger than that of IBUU04
model, degrees of model uncertainty of isospin sensitive
observables $n/p$ ratio of free nucleons and $\pi^-/\pi^+$ ratio
are thus larger than corresponding effects of nuclear symmetry
energy on these two probes \cite{LiBA1:2005}.

Shown in Fig.~\ref{fig:Fig4} is the $n/p$ ratio of free nucleons
and $\pi^-/\pi^+$ ratio as a function of kinetic energy in the
central reaction of $^{197}Au+^{197}Au$ at a beam energy of 400
MeV/nucleon simulated by the IBUU04 and the UrQMD models. For the
$n/p$ ratio of free nucleons, we can see that both models give the
same trend of n/p ratio as a function of nucleonic kinetic energy.
Again, the result of the UrQMD model is overall larger than that
of the IBUU04 model due to their different strengths of symmetry
potential as shown in the left panel of Fig.~\ref{fig:Fig2}.
Because the difference of the two symmetry potentials used in the
two models becomes larger and larger with increase of nucleon's
momentum, the difference of the values of n/p ratios of free
nucleons given by the two models also becomes larger with
nucleon's kinetic energy. From the right panel of
Fig.~\ref{fig:Fig4}, we can see that there is a cross between the
$\pi^-/\pi^+$ ratios from the UrQMD model and that from the IBUU
model. At lower kinetic energy part, the value of $\pi^-/\pi^+$
ratio from the UrQMD is much larger than that from the IBUU model,
but at high kinetic energy the value of $\pi^-/\pi^+$ ratio from
the IBUU is larger than that from the UrQMD model. This is caused
by different Coulomb action treatments in the two models. Because
most pion mesons are from resonance's decays, most pion mesons are
distributed at low energies. The result shown in the right panel
of Fig.~\ref{fig:Fig4} is consistent with the result shown in the
right panel of Fig.~\ref{fig:Fig3}. Over all, the large
model-dependence shown in Fig.~\ref{fig:Fig3} and
Fig.~\ref{fig:Fig4} should be kept into mind while comparing model
calculations with experimental data.

\subsection{nucleonic collective flows}
\begin{figure*}[htb]
\centering\emph{}
\includegraphics[width=0.99\textwidth]{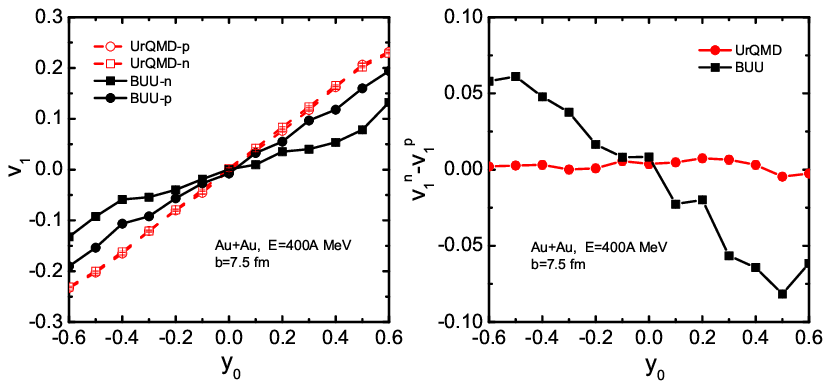}
\caption{Rapidity distributions of the neutron and proton directed
flows $<\frac{p_x}{p_t}>$ and $n-p$ directed flow difference
$v_{1}^{n}-v_{1}^p$ simulated respectively by IBUU04 model
($t_{max}$= 40 fm/c) and UrQMD model ($t_{max}$= 150 fm/c) in the
semi-central reaction of $^{197}Au+^{197}Au$ at a beam energy of
400 MeV/nucleon.} \label{fig:Fig5}
\end{figure*}
\begin{figure*}[htb]
\centering\emph{}
\includegraphics[width=0.99\textwidth]{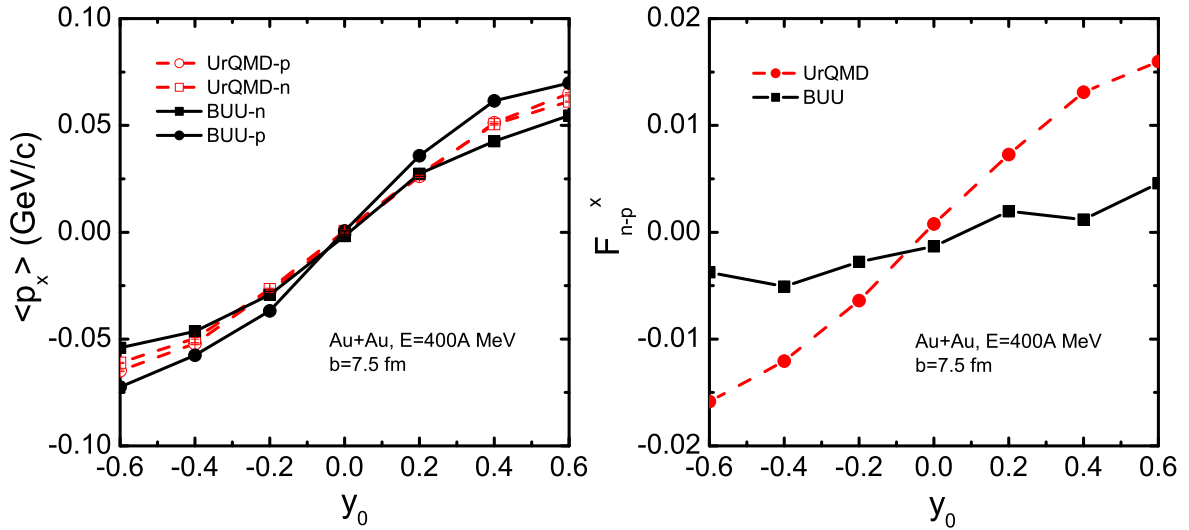}
\caption{Rapidity distributions of the transverse flow $<p_x(y)>$
for neutrons and protons and the neutron-proton differential
transverse flow $F^{x}_{n-p}$ simulated by the IBUU04 ($t_{max}$=
40 fm/c) and the UrQMD ($t_{max}$= 150 fm/c) models in the
semi-central reaction of $^{197}Au+^{197}Au$ at a beam energy of
400 MeV/nucleon.} \label{fig:Fig6}
\end{figure*}
\begin{figure*}[htb]
\centering\emph{}
\includegraphics[width=0.99\textwidth]{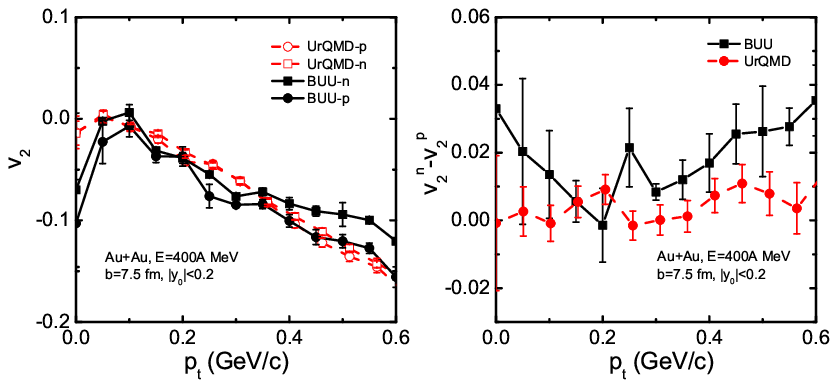}
\caption{Transverse momentum distributions of the neutron and
proton elliptic flows and the $n-p$ elliptic flow difference
$v_2^n-v_2^p$ given by the IBUU04 ($t_{max}$= 40 fm/c) and the
UrQMD ($t_{max}$= 150 fm/c) models in the semi-central reaction
$^{197}Au+^{197}Au$ at a beam energy of 400 MeV/nucleon.}
\label{fig:Fig7}
\end{figure*}
Difference of neutron and proton collective flows as a probe of
nuclear symmetry energy was first proposed by Greco \emph{et al.}
in 2003 \cite{cditoro03}. A lot of studies on such probes were
carried out in recent years
\cite{cozma11,HWolter:09,ditoro:07,WTrau:09,FengZQ:12,FengZQ1:12}.
Later on, difference of collective flows of light clusters as a
probe of nuclear symmetry energy was proposed in 2009
\cite{yongflow09}. In Fig.~\ref{fig:Fig5}, we show the reduced
rapidity distributions of neutron and proton directed flows and
$n-p$ directed flow difference $v_{1}^{n}-v_{1}^p$. Here $v_1=\
<\frac{p_x}{p_t}>$ and $v_1^n$ is the directed flow for neutrons,
$v_1^p$ is the directed flow for protons. The reduced rapidity is
$y_0=y/y_b$ and $y_b=0.8935$ is the projectile rapidity. From
Fig.~\ref{fig:Fig5} we can clearly see that the effects of isospin
on directed nucleonic flow given by the IBUU04 model is obviously
larger than that of the UrQMD model with the same symmetry energy
selection $x= 0$ ($\gamma= 0.5$). Therefore the slope of the $n-p$
directed flow $v_{1}^{n}-v_{1}^p$ given by the IBUU04 is also
evidently larger than that of the UrQMD model. This large model
dependence inevitably affects obtaining the information of
density-dependent symmetry energy from reading related
experimental data by theoretical transport model. The reason why
the nucleonic collective flow given by the IBUU04 model show large
isospin effect is that the used in-medium $NN$ cross section in
the IBUU04 model show large isospin effects as shown in the right
panel of Fig.~\ref{fig:Fig2}. The symmetry potential here in fact
does not affect the strength of nucleonic collective flow
evidently \cite{li05xsection}. Thus although the value of the
symmetry potential used in the UrQMD model is larger than that
used in the IBUU04 model, isospin effect on nucleonic collective
flow is still smaller than that calculated by the IBUU04 model
since the symmetry potential has minor effect \cite{li05xsection}.

The neutron-proton differential flow was first proposed as a probe
of nuclear symmetry energy in 2000 by Li \cite{LiBA1:2000}. This
approach utilizes constructively both the isospin fractionation
and the nuclear collective flow as well as their sensitivities to
the isospin-dependence of the nuclear equation of state. Later on,
this approach was extended to two reaction systems using different
isotopes of the same element in 2006 \cite{yonggc06}. Rapidity
dependences of the transverse flow $<p_x(y)>$ of the neutrons and
protons and the neutron-proton differential transverse flows are
shown in Fig.~\ref{fig:Fig6}. The neutron-proton differential
transverse flow is expressed as \cite{yonggc06}
\begin{eqnarray}
F_{n-p}^x(y)=\frac{N_n(y)}{N(y)}<p_x^n(y)>-\frac{N_p(y)}{N(y)}<p_x^p(y)>,
\end{eqnarray}
where $N(y)$, $N_n(y)$ and $N_p(y)$ denote the number of free
nucleons, neutrons and protons at rapidity $y$, respectively. And
$<p_x^n(y)>$ and $<p_x^p(y)>$ are the average transverse momenta
of neutrons and protons at rapidity $y$, respectively. From the
left panel of Fig.~\ref{fig:Fig6}, we can see that with the same
symmetry energy, nucleonic transverse flow given by the IBUU04
model shows large isospin effect whereas the result of the UrQMD
model does not. This is understandable since the $NN$ cross
section used in the IBUU04 model shows larger isospin effect as
discussed in Fig.~\ref{fig:Fig5}. It is noted that the slope of
neutron-proton differential flow is larger for the UrQMD model
than that for the IBUU04 model, this is understandable since the
isospin fractionation is larger for the UrQMD model as shown in
the left panel of Fig.~\ref{fig:Fig4}.

Figure~\ref{fig:Fig7} shows the transverse momentum distributions
of the neutron and proton elliptic flows and the $n-p$ elliptic
flow differences $v_2^n-v_2^p$ calculated by the IBUU04 and the
UrQMD models with the rapidity cut $\lvert{y_0}\rvert<0.2$. Here
$v_2$ is defined as $v_2=\ <\frac{p_x^2-p_y^2}{p_t^2}>$ and
$v_2^n$ is the elliptic flow for neutrons, $v_2^p$ is the elliptic
flow for protons. From Figure~\ref{fig:Fig7}, we can see that
nucleonic elliptic flows given by the two models are quite
similar. At whole transverse momenta range 0 $\sim$ 0.6 GeV/c,
there is clearly isospin effect of nucleonic elliptic flow with
the IBUU04 model due to larger isospin effect of the in-medium
$NN$ cross section used in the IBUU04 model. Such effect is less
evident with the UrQMD model. Difference of the isospin sensitive
probe $n-p$ elliptic flow difference $v_2^n-v_2^p$ given by the
two models are also clearly shown.

It is noted here that the freeze-out time of the reaction may also
affect effects of isospin of observables in heavy-ion collisions
\cite{WangYongJia2012}. In our calculations, stopping time
settings are 150 fm/c in the UrQMD model and 40 fm/c in IBUU04
model, respectively. In fact in the UrQMD model, the isospin
effect is evident before the freeze-out time of 150 fm/c.

In fact, as shown in the left panels of Figure~\ref{fig:Fig6} and
Figure~\ref{fig:Fig7}, both the IBUU04 model and the UrQMD model
give almost the same isospin-independent nucleonic collective
flows. The larger slope of $v_1=<\frac{p_x}{p_t}>$ given by the
UrQMD than that of the IBUU04 shown in the left panel of
Figure~\ref{fig:Fig5} is cause by weak squeezing out (thus smaller
$p_{x}/p_{y}$) of the UrQMD than that of the IBUU04.

\section{Conclusions and remarks}

Frequently used sensitive probes (nucleon or pion emission isospin
ratios and relative nucleonic collective flow) of nuclear symmetry
energy are simulated and compared in two different frameworks of
transport model using their ``best settings''. Sensitive probe of
n/p ratio of free nucleons is affected much by the symmetry
potential while the isospin-sensitive probes of charged
$\pi^-/\pi^+$ ratio, transverse flow and elliptic flow are
affected much by the isospin-dependent $NN$ cross section.
Different isospin effects of observables given by different
transport models originate from different forms of symmetry
potential or isospin-dependent in-medium $NN$ cross section.
Sensitive probes of nuclear symmetry energy at high densities may
suffer large uncertainties which are comparable with the effects
of nuclear symmetry energy on these probes. Therefore one must be
careful when drawing the conclusion on density-dependent nuclear
symmetry energy by reading related nuclear experiments with
transport models.

Besides improving the framework of transport model from
semi-classical transport to quantum transport \cite{daniequan}, it
would be nice to make thorough studies on the scattering cross
sections, especially, of isospin-dependent nucleon-nucleon in
medium \cite{yongpho11,lipw05,zhf0710} and symmetry potential of
nucleon in asymmetric matter \cite{chenprc12}. And also some
unknown nucleon-nucleon interaction such as the tensor force
induced isospin-dependence of short-range nucleon-nucleon
correlation \cite{xuli} and spin-orbit potential \cite{xuli13} may
also affect isospin sensitive observables in heavy-ion collisions.
Therefore searching for probes that insensitive to the
uncertainties of model inputs or large sensitive probe such as
possible $\eta$ production in heavy-ion collisions \cite{eta13}
are always useful for the study of high-density nuclear symmetry
energy.

\section{Acknowledgements}
We acknowledge support by the computing server C3S2 in Huzhou
Teachers College. The work is supported by the National Natural
Science Foundation of China (Grant Nos. 11175219,10905021,
10979023, 11175074, 11075215, 11275052), the 973 Program of
China (Grant No. 2013CB834405), the Knowledge Innovation Project
(Grant No. KJCX2-EW-N01) of Chinese Academy of Sciences,
the Chinese Academy of Sciences via a grant
for visiting senior international scientists (Grant No.
2009J2-26), the Zhejiang Provincial Natural Science Foundation of
China (No. Y6090210), the Qian-Jiang Talents Project of Zhejiang
Province (No. 2010R10102), and the National Key Basic Research
Program of China (No. 2013CB834400), the Project of Knowledge
Innovation Pro- gram (PKIP) of Chinese Academy of Sciences (Grant
No. KJCX2.YW.W10), and the CAS/SAFEA International Partnership
Program for Creative Research Teams (Grant No. CXTD-J2005-1).

\end{document}